\documentclass{elsart}
\usepackage{amssymb,epsfig}
 
\def\be{\begin{equation}}
\def\ee{\end{equation}}
\def\bq{\begin{eqnarray}}
\def\eq{\end{eqnarray}}
 
\begin{document}
\begin{frontmatter}

\title{From Quark Stars to Hybrid Stars}
\author{Alessandro Drago$^a$ and Andrea Lavagno$^b$}
\address{$^a$Dipartimento di Fisica, Universit{\`a} di Ferrara 
and INFN, Sezione di Ferrara, 44100 Ferrara, Italy}
\address{$^b$Dipartimento di Fisica, Politecnico di Torino 
and INFN, Sezione di Torino, 10129 Torino, Italy}

\date{\today }
\maketitle

\begin{abstract}
We show the possible existence of compact stars having a surface
composed of a mixed phase of quarks and hadrons. This scenario
can be realized both for self-bound stars, satisfying the 
so-called Witten-Bodmer hypothesis, and for gravitationally
bound stars. This class of solutions of the 
Tolman-Oppenheimer-Volkoff equation can be obtained in all the models
we discuss, within a physically acceptable
range of values of the model parameters.


\end{abstract}
\end{frontmatter}

Recently several analysis of observational data have emphasized 
the possible existence of compact stars having very small radii,
of the order of 9 kilometers or less \cite{li99}.  
To account theoretically for the existence
of such a star, one would need to use of very soft Equations of State (EOSs),
that can be obtained, in principle, introducing kaon condensation 
\cite{kaplan,thorsson,glen1}, 
or hyperonic degrees of freedom \cite{glen2,schaffner,schaffner2}. 
Another possibility
is to consider quark stars, where the density of the surface
is above nuclear matter saturation density. 
These objects have been
obtained till now assuming the validity of the 
hypothesis introduced independently by 
Bodmer \cite {bodmer} and Witten \cite{witten}, which states that 
the true and absolute ground state of the strong interaction is a 
state of deconfined quark matter.
The latter consists of an approximately equal number 
of up, down and strange quarks having energy per baryon $E/A$ smaller
than that of iron ($E_{Fe}\approx 930$ MeV), 
at zero temperature and pressure \cite{farhi}. 
Stars entirely composed of matter in such ultra-stable state are self-bound 
\cite{alcock,haensel} and could rotate with a 
period well below one millisecond \cite{glen_rot,madsen}.

A possible problem in interpreting neutron stars as strange quark stars is
the difficulty of having glitches in the latter, since the most
widely accepted model for the glitches require the existence
of a crystallized phase trapping magnetic flux tubes. It is not
clear whether such a structure can be obtained in quark stars, 
although the presence of a color superconductive phase could
open this possibility \cite{raja}.

The aim of this letter is to show that
it is possible to find a minimum of the energy per baryon number 
at finite density, in the mixed quark-hadron phase. 
This minimum corresponds to an energy 
which can be either above or below $E_{Fe}$. 
Stars builded using this EOS will be
gravitationally bound, if the minimum is above $E_{Fe}$,
or self-bound stars if the minimum is below $E_{Fe}$. In both cases
the surface of the star is made of mixed quark-hadron phase. 
To our knowledge, this possibility has never been
investigated, because in general the so-called quarks stars have
been studied using the quark EOS alone, without matching it
with the hadronic EOS. In this paper, on the contrary, we will study in detail
the transition from a low density EOS (hadrons) to a high density one
(quarks), using the Gibbs conditions.  

The EOS appropriate to the description of a compact star has to
satisfy $\beta$-stability conditions. This implies the
existence of two conserved charges, the baryonic charge and the electric 
charge.
When the Gibbs conditions are applied in presence of more
than a single conserved charge, the technique developed by 
Glendenning has to be adopted \cite{glen4,glen5,schaffner3}. 
Using that technique,
the pressure need not to be constant in the mixed phase, what will
be crucial in our investigation.

The conservation of the  baryon
(B) and the electric (C) charge can be written as follows: 
\bq 
\rho_B&=&(1-\chi)\rho_B^h+\chi\rho_B^q\\
\rho_C&=&(1-\chi)\rho_C^h+\chi\rho_C^q+\rho_e+\rho_\mu=0\,\nonumber .
\eq
Here $\chi$ is the fraction of matter in the quark phase. The 
superscripts $h$ and $q$ label the density in the hadronic and in the
quark phase, respectively. The electron ($\rho_e$) and the muon
($\rho_\mu$) charge densities 
contribute to make the total electric charge equal to zero.

The equations of chemical equilibrium under $\beta$-decay and
deconfinement are the following:
\bq
\mu_n-\mu_p=\mu_e \ \  &,& \quad
\mu_n-\mu_p=\mu_\mu \nonumber\\
2\mu_d+\mu_u=\mu_n\ \  &,& \quad\mu_u-\mu_d=\mu_p-\mu_n\nonumber\\
\mu_s&=&\mu_d\,.
\eq
The usual condition for mechanical equilibrium, {\it i.e.}
the equality of the pressure in the two phases, reads:
\be
P^h=P^q \,.
\ee
The previous equations have to be solved together with the
field's equations for the adopted hadronic and quark models.
Concerning the hadronic phase, we have used the relativistic non-linear
Walecka-type models of Glendenning-Moszkowsky 
(GM1, GM2, GM3) \cite{glen2} and Bodmer (B91) \cite{b91}. 
For the quark phase we have considered the MIT bag model 
at first order in the strong coupling constant $\alpha_s$ 
\cite{farhi} and the
Color Dielectric Model (CDM) \cite{pirner,birse}. In the latter, quarks develop
a density dependent constituent mass through their interaction
with a scalar field $\kappa$ representing a multi-gluon state. 
The Lagrangian of the CDM reads:
\begin{eqnarray}
     {L} &=& i\bar \psi \gamma^{\mu}\partial_{\mu} \psi 
 +{1\over 2}{\left(\partial_\mu\sigma \right)}^2
     +{1\over 2}{\left(\partial_\mu\vec\pi\right)}^2
     -U\left(\sigma ,\vec\pi\right)   \nonumber\\
     &+&\!\!\!\sum_{f=u,d} {g_f\over f_\pi \kappa} \, \bar \psi_f\left(\sigma
     +i\gamma_5\vec\tau\cdot\vec\pi\right) \psi_f 
     +{g_s \over \kappa} \, \bar \psi_s \psi_s       \nonumber
     \\
      &+&{1\over 2}{\left(\partial_\mu\kappa\right)}^2
      -{1\over 2}{M}^2\kappa^2
\end{eqnarray}
where $U(\sigma ,\vec\pi)$ is the ``mexican-hat'' potential, as in
Ref.\ \cite{neuber}. 

The coupling constants are given by $g_{u,d}=g (f_{\pi}\pm \xi_3)$
and $g_s= g(2 f_K -f_{\pi})$, where $f_{\pi}=93$ MeV and $f_{K}=113$ MeV 
are the pion and the kaon 
decay constants, respectively, and $\xi_3=f_{K^\pm}-f_{K^0}=-0.75$ MeV. 
These coupling constants depend on a single parameter $g$.
Confinement is obtained {\it via} the effective quark masses
$m_{u,d}=-g_{u,d} \bar\sigma/(\bar\kappa f_\pi)$ and 
$m_s=g_s / \bar\kappa$,
which diverge outside the nucleon.
Working at mean-field level, the only free parameter is actually the
product $G=\sqrt{g {M}}$. In this model the last term of the
lagrangian plays a role similar to the vacuum pressure constant $B$
of the MIT bag model. At variance with the latter, in the CDM
the vacuum pressure is density dependent.

\begin{figure}
\parbox{6cm}{
\scalebox{0.6}{
\includegraphics*[-5,410][550,800]{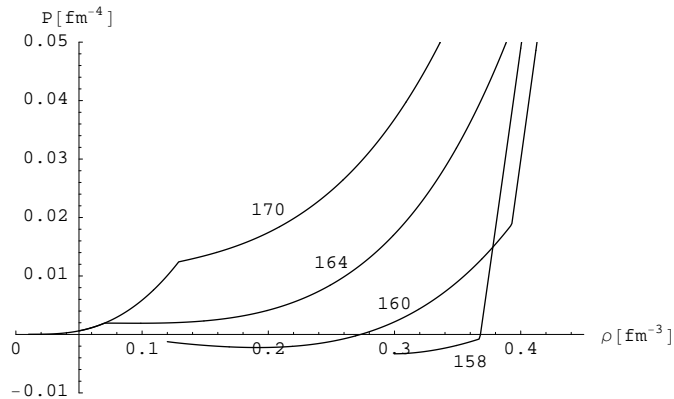}}
}

\parbox{14cm}{
\caption{The pressure as a function of the baryon density for various 
values of the MIT-bag constant $B^{1/4}$ [MeV], with $\alpha_s=0$, 
$m_u=m_d=0$, $m_s=100$ MeV and the GM3 hadronic model \cite{glen2}.}}
\vspace{1cm}
\end{figure}

Let us discuss our results about the EOS in the mixed hadron-quark phase 
applying the above discussed Gibbs conditions. In Fig. 1, we report 
the pressure as a function of the baryon density for different 
values of the MIT-bag constant $B^{1/4}$ [MeV], with $\alpha_s=0$, 
$m_u=m_d=0$, $m_s=100$ MeV 
and the GM3 hadronic model \cite{glen2}.
The figure shows that decreasing the bag constant from 
the value $B^{1/4}=170$ MeV, corresponding to a typical   
hybrid star EOS  
(composed by a central quark phase, an intermediate 
mixed phase and an hadronic crust at low density),  
to the value $B^{1/4}=158$ MeV, corresponding to 
a pure quark star (with a nonzero density at the surface), 
we have a region where the pressure decreases by increasing the 
baryon density (instability condition) 
and, for some values of $B^{1/4}$, we have 
an EOS which exhibits a double zero of the pressure. 
In particular, in 
the curve with $B^{1/4}=160$ MeV, the second zero of the pressure 
corresponds to a second minimum of the energy per baryon number 
at finite baryon density ($\rho=0.274$ fm$^{-3}$) 
in the quark-hadron mixed phase 
($\chi=0.55$).\footnote{For consistency, we have checked that
for all the hadron/quark models we have considered, 
the phase transition from the hadronic 
matter to the hadron-quark mixed phase in (non-strange) 
symmetric nuclear matter occurs 
at baryon densities much larger than the ordinary nuclear matter density.}  
There is therefore a range of quark model parameters for which 
it is possible to realize stable 
hybrid stars with a sharp surface composed by hadrons and quarks. 

The previous analysis is based on the use of Gibbs conditions,
which for $\beta$-stable matter are not equivalent to 
the so-called Maxwell construction. The latter would actually not
be possible when the minimum of the quark EOS corresponds
to an energy per baryon number smaller than the nucleon mass.
Gibbs construction is at the contrary still possible, and 
reduces the energy of the system in the mixed phase. 
   
It is interesting to note that a very similar behavior of EOS 
has been recently obtained in Ref. \cite{schaffner2} 
studying phase transition to hyperonic matter in neutron stars. Increasing the 
strengths of the hyperon-hyperon interactions, the authors find a second 
minimum of the energy per baryon (i.e. a second zero of the pressure) 
at a large strangeness fraction. 
Such minimum can be deeper than ordinary matter 
(absolutely stable strange hadronic matter and, consequently, self-bound 
neutron stars) or not (gravitationally bound stars). 
Analogous results are obtained by us in the context of the quark-hadron 
phase transition.

\begin{figure}
\parbox{6cm}{
\scalebox{0.6}{
\includegraphics*[-70,70][430,750]{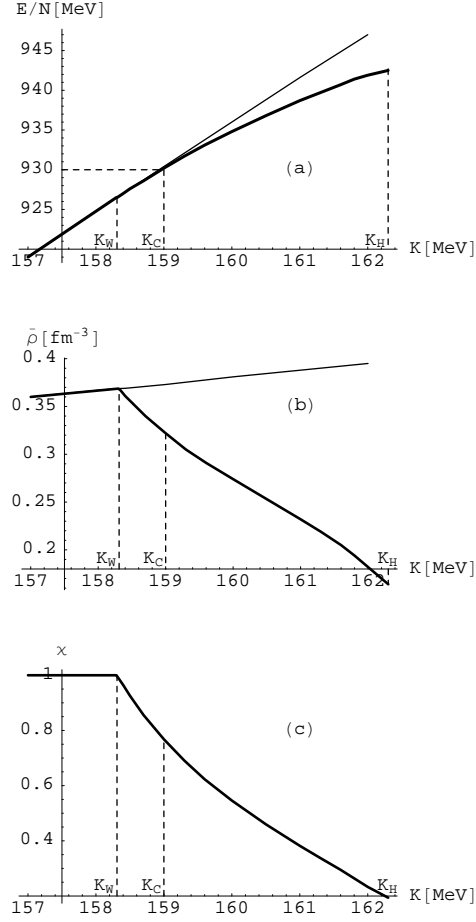}}
}

\parbox{14cm}{
\caption{(a) The energy per baryon number $E/N$ at its second minimum 
as a function of $K\equiv B^{1/4}$ [MeV];  
(b) the corresponding baryon density $\overline{\rho}$;   
(c) the associate fraction of quark matter $\chi$. 
In Fig. 2a and 2b the thin line corresponds to a pure 
quark EOS 
(MIT bag model with $\alpha_s=0$, $m_u=m_d=0$, $m_s=100$ MeV) while the thick 
line corresponds to matching the quark and the hadronic EOS using 
Gibbs conditions. }}
\vspace{1cm}
\end{figure}

To understand better our results, we show in Fig. 2: (a) the 
values of the second minimum 
of the energy per baryon number $E/N$; (b) the corresponding 
baryon density $\overline{\rho}$ and (c) the associate 
fraction of quark matter $\chi$, for different 
$K_i\equiv B_i^{1/4}$ [MeV]. 
In Fig. 2a and 2b the thin line represents the results 
obtained using a pure quark EOS 
(MIT model, with $\alpha_s=0$, $m_u=m_d=0$, $m_s=100$ MeV, 
the dependence of our results on $\alpha_s$ and $m_s$ will be discussed below) 
while the thick line corresponds to matching the quark and the hadronic EOS 
using the Gibbs conditions. 
For values of the bag constant smaller 
than $K_W$, we have self-bound purely quark 
stars: the second minimum of the energy per baryon number 
lies in the pure quark phase. 
By increasing $K$, we have stars with an increasing fraction of 
hadronic matter at the surface ($\chi<1$). 
In the range between $K_W$ and $K_C$ the energy 
per particle is less then $930$ MeV and 
the Bodmer-Witten condition is satisfied: we have therefore 
self-bound stars with a surface composed 
by a quark-hadron mixed phase. 
For values $K_C<K<K_H$, we have gravitationally bound stars 
with a sharp edge, while, for $K>K_H$, we obtain 
hybrid stars with a density profile vanishing to zero. 

Since we have introduced in the previous paragraph a class
of not self-bound stars, it is natural to address the question
of their stability. In particular it is important to check 
that low-mass stars are stable against ``evaporation'' into iron nuclei.
A sufficient condition is the following:

\be
\frac{\partial M}{\partial N}< E_{Fe},
\ee
where $M$ is the gravitational mass of the star and $N$ is the baryon
number. We have checked numerically that,
in all models we have considered, stars having a radius
larger than $4$ km (or equivalently $M\gtrsim 0.1 M_\odot$) are stable.
Stars for which observational constraints on the 
$M$--$R$ relation has
been discussed in the literature should have a mass larger than
$1 M_\odot$. Therefore we will not discuss in detail the low mass region.

\begin{table}[ht]
\begin{center}
\begin{tabular}{cccc}
\hline
\hline
Models & $K_{Min}$ & $K_W$, $E/N$ & 
$K_H$, $E/N$, $\chi$ \\
\hline
GM1-MIT, $\alpha_s=0$, $m_s=100$ MeV & $145$
& $158.4$, $927$ & $162.5$, $943$, $0.16$ \\  
GM3-MIT, $\alpha_s=0$, $m_s=100$ MeV & $145$
& $158.3$, $927$ & $162.3$, $943$, $0.17$ \\  
B91 -MIT, $\alpha_s=0$, $m_s=100$ MeV & $145$
& $159.2$, $931$ & $161.8$, $942$, $0.13$ \\  
\hline
GM1-MIT, $\alpha_s=0$, $m_s=150$ MeV & $145$
& $155.1$, $931$ & $158.1$, $943$, $0.20$ \\
GM3-MIT, $\alpha_s=0$, $m_s=150$ MeV & $145$
& $155.0$, $930$ & $157.9$, $943$, $0.18$ \\
B91 -MIT, $\alpha_s=0$, $m_s=150$ MeV & $145$
& $155.9$, $935$ & $157.4$, $942$, $0.22$ \\  
\hline
GM1-MIT, $\alpha_s=0.3$, $m_s=100$ MeV & $137.5$
& $150.7$, $929$ & $154.2$, $943$, $0.25$ \\  
GM3-MIT, $\alpha_s=0.3$, $m_s=100$ MeV & $137.5$
& $150.6$, $929$ & $154.0$, $943$, $0.23$ \\  
B91 -MIT, $\alpha_s=0.3$, $m_s=100$ MeV & $137.5$
& $151.4$, $934$ & $153.4$, $943$, $0.23$ \\  
\hline
GM1-MIT, $\alpha_s=0.3$, $m_s=150$ MeV & $137.5$
& $147.4$, $932$ & $150.2$, $944$, $0.25$ \\
GM3-MIT, $\alpha_s=0.3$, $m_s=150$ MeV & $137.5$
& $147.4$, $931$ & $150.0$, $943$, $0.27$ \\
B91 -MIT, $\alpha_s=0.3$, $m_s=150$ MeV & $137.5$
& $148.2$, $936$ & $149.5$, $942$, $0.27$ \\
\hline  
GM3-MIT, $\alpha_s=0$, $m_s=200$ MeV & $145$
& $151.6$, $936$ & $153.1$, $943$, $0.31$ \\
GM3-MIT, $\alpha_s=0.3$, $m_s=200$ MeV & $137.5$
& $143.9$, $935$ & $145.6$, $943$, $0.29$ \\
GM3-MIT, $\alpha_s=0.6$, $m_s=200$ MeV & $128.1$
& $134.8$, $934$ & $136.7$, $943$, $0.33$ \\
\hline
GM1-CDM & $19.8$
& $21.9$, $932$ & $23.2$, $944$, $0.19$ \\  
GM3-CDM & $19.8$
& $21.9$, $932$ & $23.1$, $942$, $0.22$ \\  
B91 -CDM & $19.8$
& $22.3$, $936$ & $22.8$, $942$, $0.29$ \\  

\hline
\hline
\end{tabular}
\vspace{0.5cm}
\caption{Critical values $K_{Min}$, 
$K_W$ and $K_H$ of the quark models parameter $K$
($K\equiv B^{1/4}$ [MeV], for MIT-bag model and 
$K\equiv g$ [MeV], for CDM model). 
$K_{Min}$ is the minimum $K$ for 
which the energy of non-strange quark matter is above $E_{Fe}$. For $K<K_W$  
we have pure quark stars, for $K_W<K<K_H$ we have hybrid stars 
with a surface made of quark-hadron mixed phase, for $K>K_H$ we have
normal hybrid stars with hadronic matter density vanishing at the surface. 
$E/N$ [MeV] is the energy per baryon number
at the minimum
(both for $K=K_W$ and for $K=K_H$),
$\chi$ is the fraction of mixed phase made of quark matter.}
\end{center}
\label{}
\end{table}

In Table 1, we report the results for different 
hadronic \cite{glen2,b91} and quark models, like the
MIT-bag model ($K_i\equiv B_i^{1/4}$ [MeV]) 
and CDM model ($K_i\equiv g_i$ [MeV] and  
${M}=1.7$ GeV). $K_{Min}$ is the minimum $K$ for 
which non-strange quark matter is unbound (the energy per baryon 
of non-strange 
quark matter must exceed the lowest energy per baryon in nuclei, i.e., 
about $930$ MeV for iron). We observe, from the third column 
of the table ($K_W$ and $E/N$), that only for specific values of 
quark model parameters it is possible to realize absolutely 
stable self-bound matter in mixed phase ($E/N \leq 930$ MeV). 
For large values of $\alpha_s$ and $m_s$ 
the Bodmer-Witten hypothesis is satisfied only in pure 
quark matter (in other words: $K_C<K_W$). 
Let us note that the values of the quark model parameters used 
in the present calculation are close to
the values adopted in hadronic calculations (for example,  
the coupling  $g\approx 22$ MeV in the CDM is very near to the value 
$g=23$ MeV used to reproduce the experimental mass and 
radius of the nucleon \cite{neuber}).

\begin{figure}
\parbox{6cm}{
\scalebox{0.6}{
\includegraphics*[-5,470][550,800]{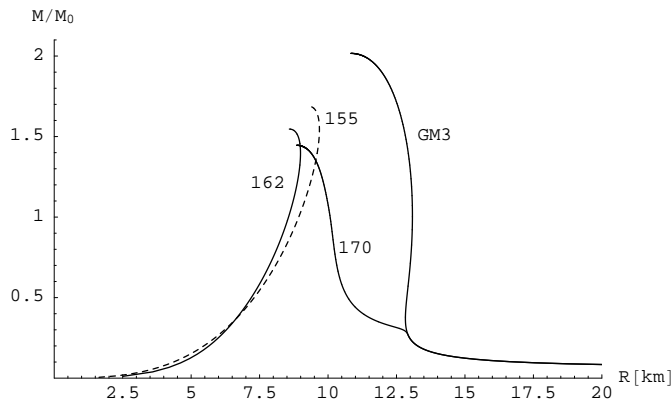}}
}

\parbox{14cm}{
\caption{The mass-radius relation for different EOSs. 
The curves GM3 and $B^{1/4}=170$ MeV are ordinary neutron stars with 
a vanishing hadronic density at the surface. 
The solutions with $B^{1/4}=155$ MeV are pure quark stars while  
$B^{1/4}=162$ MeV corresponds to hybrid stars with a quark matter core and a 
surface composed of mixed phase ($\chi=0.23$). }}
\vspace{1cm}
\end{figure}

The stable solutions of the Tolman-Oppenheimer-Volkoff equation are 
shown in Figs. 3 (mass-radius relation) and 4 
(energy density profile at fixed gravitational mass $M=1.4 \, M_\odot$). 
In both figures the 
curve labeled with GM3 corresponds to the solution obtained using 
the hadronic 
EOS alone (no phase transition), 
while the others are labeled with the adopted value of
$B^{1/4}$ [MeV] of the 
MIT-bag model ($\alpha_s=0$, $m_s=100$ MeV). 
The curves GM3 and $B^{1/4}=170$ MeV are ordinary neutron stars with hadronic 
density vanishing at the surface 
(for subnuclear densities we have employed the 
Baym-Pethick-Sutherland EOS \cite{bps}, which describes the crust of the 
neutron star). 
The other two curves are crustless stars 
with a non-zero density at the surface. 
The solution with $B^{1/4}=155$ MeV corresponds to 
pure quark stars while the one with $B^{1/4}=162$ MeV
represents hybrid stars with a central quark matter core and a 
surface composed of mixed phase ($\chi=0.23$). 
Similar behaviors are obtained by using the CDM.

\begin{figure}
\parbox{6cm}{
\scalebox{0.6}{
\includegraphics*[-35,470][550,800]{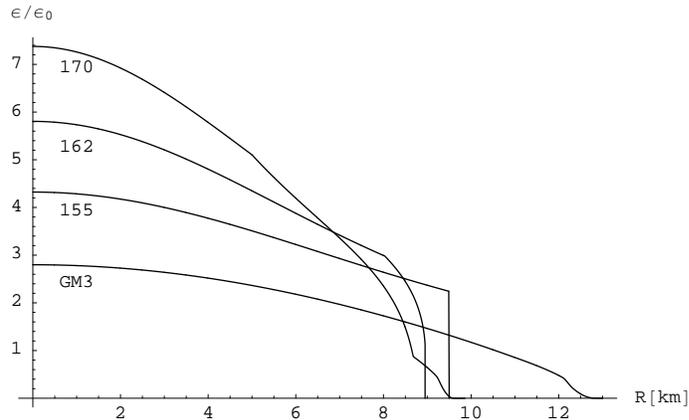}}
}

\parbox{14cm}{
\caption{Energy density profile (in units of the nuclear matter energy density
at saturation $\epsilon_0$) 
at fixed stellar gravitational mass $M=1.4 \, M_\odot$. 
Labels as in Fig. 3. }}
\vspace{1cm}
\end{figure}

In the present letter, we have not considered the possibility 
of having a thin 
crust of nuclear material, suspended from contact with the 
high density mixed phase by an electric dipole field \cite{alcock,glen5}. 
The introduction of this thin crust would 
modify significantly the $M$--$R$ 
relationship only for extremely small values of the mass 
($M\approx 0.1 M_\odot$) \cite{schaffner2}. As discussed above, we are not 
interested in this range of low mass stars.

The main result of our work is the indication of the possibility to have 
stars whose surface is composed of a mixed phase of quarks and hadrons. 
For certain ranges of model parameters, these solutions of the 
Tolman-Oppenheimer-Volkoff equation can be obtained without the need to 
satisfy the Bodmer-Witten hypothesis and corresponds to
gravitationally bound stars. In other parameter ranges, 
it is possible to satisfy Bodmer-Witten hypothesis in
mixed phase and self bound stars are obtained. We stress again that
the present analysis could not be done using the Maxwell construction
instead of Gibbs conditions, since no Maxwell construction
would be possible for a quarks' EOS having a minimum of the energy
per baryon number below the mass of the nucleon. 

The existence of mixed phase at the surface of a compact 
star could be crucial  
in the explanation of the glitches, since in the mixed phase 
a crystalline structure can develop once the surface tension is taken 
into account. Moreover a large fraction of hadronic matter in the surface 
of the star seems to be needed at the light of recent works on r-mode 
instability \cite{madsen} and on the cooling of quark stars 
\cite{blaschke,lattimer}.

It is a pleasure to thank V.~Barone for a careful reading of the
manuscript.
This work is supported in part by MURST.


\begin{thebibliography}{}
\bibitem{li99}
X.-D. Li {\it et al.}, Phys. Rev. Lett. 83 (1999) 3776; 
M. Dey {\it et al.}, Phys. Lett. B 438 (1998) 123.
\bibitem{kaplan}
D.B. Kaplan and A.E. Nelson, Phys. Lett. B 175 (1986) 57; 179 (1986) 409. 
\bibitem{thorsson}
V. Thorsson, M. Prakash and J.M. Lattimer, Nucl. Phys. A 572 (1994) 693. 
\bibitem{glen1}
N.K. Glendenning, J. Schaffner-Bielich, Phys. Rev. Lett. 81 (1999) 4564; 
Phys.  Rev. C 60, 025803.
\bibitem{glen2}
N.K. Glendenning, S.A. Moszkowski, Phys. Rev. Lett. 67 (1991) 2414. 
\bibitem{schaffner}
J. Schaffner, I.N. Mishustin, Phys. Rev. C 53 (1996) 1416. 
\bibitem{schaffner2}
J. Schaffner-Bielich, M. Hanauske, H. St\"ocker, W. Greiner, astro-ph/0005490.
\bibitem{bodmer}
A.R. Bodmer, Phys. Rev. D 4 (1971) 1601. 
\bibitem{witten}
E. Witten, Phys. Rev. D 30 (1984) 272. 
\bibitem{farhi}
E. Farhi and R.L. Jaffe, Phys. Rev. D 30 (1984) 2379. 
\bibitem{alcock}
C. Alcock, E. Fahri, A. Olinto, ApJ. 310 (1986) 261.
\bibitem{haensel}
P. Haensel, J.L. Zdunik, R. Schaeffer, Astron. Astrophys. 160 (1986) 121.
\bibitem{glen_rot}
N.K. Glendenning, Mod. Phys. Lett. A 5 (1990) 2197.
\bibitem{madsen}
J. Madsen, Phys. Rev. Lett. 81 (1998) 3311; Phys. Rev. Lett. 85 (2000) 10.
\bibitem{raja}
M. Alford, J. Bowers, K. Rajagopal, 
J. Phys. G 27 (2001) 541; 
hep-ph/0008208, Phys. Rev. D, in press.
\bibitem{glen4}
N.K. Glendenning, Phys. Rev. D 46 (1992) 1274;
\bibitem{glen5}
N.K. Glendenning, {\it Compact Stars}, 1997 Springer-Verlag, New York.
\bibitem{schaffner3}
K. Schertler, C. Greiner, J. Schaffner-Bielich, M.H. Thoma, 
Nucl. Phys. A 677 (2000) 463. 
\bibitem{b91}
A.R. Bodmer, Nucl. Phys. A 526 (1991) 703.
\bibitem{pirner}
H.J. Pirner, Prog. Part. Nucl. Phys. 29 (1992) 33.
\bibitem{birse}
M.C. Birse, Prog. Part. Nucl. Phys. 25 (1990) 1; 
J.A. McGoven, Nucl. Phys. A 533 (1991) 553.
\bibitem{neuber}
T. Neuber, M. Fiolhais, K. Goeke, J.N. Urbano, Nucl. Phys. A 560 (1993) 509. 
\bibitem{bps}
G. Baym, C.J. Pethick, P. Sutherland, ApJ. 170 (1971) 299; 
G. Baym, H.A. Bethe, C.P. Pethick, Nucl. Phys. A 175 (1971) 225.
\bibitem{blaschke}
D. Blaschke, T. Kl\"ahn, D.N. Voskresensky, ApJ. 533 (2000) 406.
\bibitem{lattimer}
A.W. Steiner, M. Prakash, J.M. Lattimer, Phys. Lett. B 486 (2000) 239 and 
astro-ph/0101566. 


\end{thebibliography}
\end{document}